\documentclass[letterpaper]{article} 
\usepackage{aaai2026}  
\usepackage{times}  
\usepackage{helvet}  
\usepackage{courier}  
\usepackage[hyphens]{url}  
\usepackage{graphicx} 
\usepackage{natbib}  
\usepackage{caption} 
\usepackage{booktabs} 
\usepackage{multirow}
\usepackage{amsmath}
\usepackage{adjustbox}
\usepackage{subcaption}  
\usepackage{algorithm}
\usepackage{algorithmic}

\usepackage{amsfonts}
\usepackage{newfloat}
\usepackage{listings}
\DeclareCaptionStyle{ruled}{labelfont=normalfont,labelsep=colon,strut=off} 
\floatstyle{ruled}
\newfloat{listing}{tb}{lst}{}
\floatname{listing}{Listing}
\nocopyright
\title{KALL-E: Autoregressive Speech Synthesis with Next-Distribution Prediction
}
\author{
    Kangxiang Xia\textsuperscript{\rm 1}\thanks{These authors contributed equally as first authors.},
    Xinfa Zhu\textsuperscript{\rm 1}\footnotemark[1],
    Jixun Yao\textsuperscript{\rm 1},
    Wenjie Tian\textsuperscript{\rm 1},
    Wenhao Li\textsuperscript{\rm 1},
    Lei Xie\textsuperscript{\rm 1}\thanks{Corresponding author.}
}

\affiliations{
    \textsuperscript{\rm 1}Audio, Speech and Language Processing Group (ASLP@NPU),\\
Northwestern Polytechnical University, Xi’an\\
    xkx@mail.nwpu.edu.cn, lxie@nwpu.edu.cn\\
    \vspace{0.5em} 
    \centering
    \textbf{Project page:} \url{https://github.com/xkx-hub/KALL-E}
}

\usepackage{bibentry}

\begin{document}

\maketitle

\begin{abstract}
We introduce KALL-E, a novel autoregressive (AR) language model for text-to-speech (TTS) synthesis that operates by predicting the next distribution of continuous speech frames. Unlike existing methods, KALL-E directly models the continuous speech distribution conditioned on text, eliminating the need for any diffusion-based components. Specifically, we utilize a Flow-VAE to extract a continuous latent speech representation from waveforms, instead of relying on discrete speech tokens. A single AR Transformer is then trained to predict these continuous speech distributions from text, optimizing a Kullback–Leibler divergence loss as its objective. Experimental results demonstrate that KALL-E achieves superior speech synthesis quality and can even adapt to a target speaker from just a single sample. Importantly, KALL-E provides a more direct and effective approach for utilizing continuous speech representations in TTS.

\end{abstract}

\vspace{-6pt}
\section{Introduction}
Large language models (LLMs), known for their impressive zero-shot and in-context learning abilities, have unified the research paradigm in natural language processing. Inspired by the success of text-based LLMs~\cite{chatgpt4}, the next-token prediction framework has been extended to tasks in other modalities. Text-to-speech (TTS), for instance, has recently gained popularity by being framed as a next-token prediction problem~\cite{valle}~\cite{valle2}. In this approach, speech signals are first tokenized into sequences of discrete units using vector quantization~\cite{encodec}~\cite{soundstream}. A decoder-only language model is then trained on these acoustic tokens. However, unlike naturally discrete text data, speech discretization relies on complex quantization techniques and presents additional challenges.

Despite significant efforts to enhance discrete speech tokenizers~\cite{speechtokenizer}~\cite{singlecodec}~\cite{xcodec}~\cite{wavtokenizer}, serious issues remain.
First, discrete speech tokens may lose certain information during tokenization~\cite{melle}, even if this loss is imperceptible to humans or specific models. Similar issues have been observed in some vision studies~\cite{givt}~\cite{image-vae}, where the reconstruction quality of quantized tokens typically falls short of their continuous-valued counterparts.
Second, many discrete speech tokens tend to have a high frame rate, resulting in strong similarities between tokens over short periods. This can cause language models to generate long stretches of silence or continuous noise, leading to hallucinations~\cite{ELLA-V}.
Some tokenizers use multiple discrete tokens per speech frame to capture richer acoustic details~\cite{dac}, but this greatly increases the training complexity of the language model. On the other hand, recently single-codebook codecs preserve less acoustic information and depend on a more powerful generation module to enhance details~\cite{cosyvoice}. This multistep process reduces inference efficiency and raises computational costs.
\begin{figure}[tbp]
    \centering
    \includegraphics[width=\linewidth]{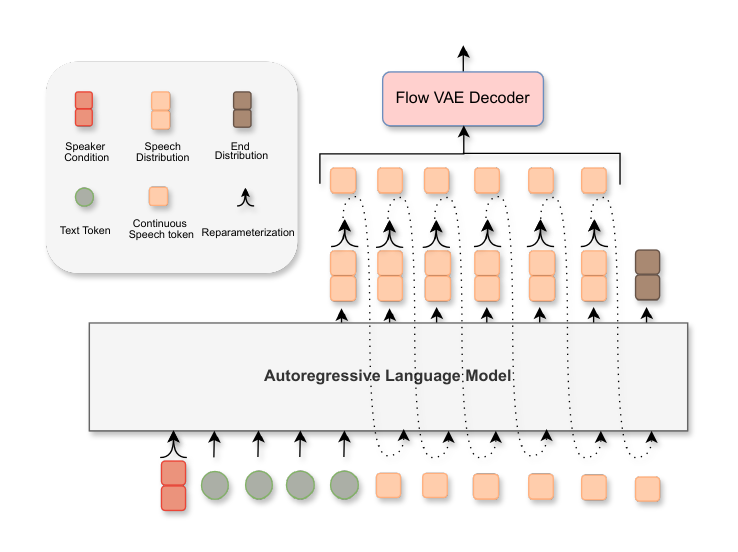}
    \caption{Overview of KALL-E. It predicts the continuous speech distribution frame by frame.}
    \label{fig:architecture}
    \vspace{-20pt}
\end{figure}

Recent studies~\cite{melle}~\cite{felle}~\cite{ditar} have explored continuous speech representations within autoregressive language modeling frameworks to address the limitations of discrete speech tokenization. 
However, modeling these continuous representations introduces a range of challenges.
Traditional regression-based loss functions used in MELLE~\cite{melle}, such as mean absolute error (MAE) and mean squared error (MSE), rely on overly simplified distributional assumptions. These assumptions often fail to capture the multimodal and complex nature of training data, leading to predictions that are ambiguous, oversimplified, or overly averaged. Some studies have introduced additional processing architecture after the language model to enhance the capacity of continuous representations modeling. However, this often leads to a complex model structure. Moreover, relying on post-processing networks to predict continuous representations can undermine the language model’s overall modeling capacity.

A further limitation of existing approaches lies in the failure to fully exploit the ability of continuous representation to carry information from speech almost without loss.
Continuous representation avoids information loss caused by quantization, but the frame rate of continuous representation modeled in many works is even several times that of discrete tokens.
For LM-based TTS, high frame rate modeling objects will affect modeling stability and limit the maximum modeling time. On the other hand, the increase in sequence length will increase the amount of calculation for autoregressive LM geometrically, reducing the inference speed.

To solve the above problems, we propose KALL-E, a novel AR speech synthesis approach with next-distribution prediction.
First, KALL-E uses the Kullback-Leibler (KL) divergence loss for next-distribution prediction instead of the traditional cross-entropy loss. This method models speech features in continuous space, avoiding a series of information loss issues caused by quantization.
Second, we explore a novel representation that is more suitable for continuous autoregressive modeling. By introducing flow in VAE, the extracted continuous features no longer obey the strict Gaussian prior. This brings more diverse generation results and more robust modeling.
To fully leverage the high information density of continuous representations, KALL-E operates at a low frame rate of 12.5 Hz, significantly boosting inference speed.
In addition, the test-time training (TTT) method enables KALL-E to further learn the target distribution from a single test data sample, which fully uses the advantages of high information density of continuous representation and further improves the model capability. 
For our experiments, we use the Emilia dataset as the primary training data. We evaluate KALL-E on Seed TTS eval~\cite{seedtts} and compare it with several popular TTS systems, including Seed-TTS~\cite{seedtts}, FireRedTTS~\cite{fireredtts}, CosyVoice~\cite{cosyvoice}, and Llasa~\cite{llasa}. 
Our experimental results demonstrate that KALL-E achieves impressive performance in both speed and accuracy. Specifically, we observe a substantial reduction in word error rate (WER) compared to previous models, while maintaining high naturalness in synthesis. Moreover, KALL-E excels in generating expressive, context-aware speech, showcasing its ability to handle complex linguistic and emotional features with ease. Demos can be found in the supplementary material.

Our contributions can be summarized as follows:

\begin{itemize}
    \item We propose KALL-E, a novel AR speech synthesis approach with next-distribution prediction. 
    \item We explore Flow-VAE, which not only reconstructs speech signals with high quality but also captures a richer acoustic distribution that is more suitable for autoregressive LM modeling.
    \item Remarkably, KALLE-E synthesizes speech at a low frame rate, which improves the inference speed, and achieves the lowest word error rate.
    \item We open-source our model and code to further advance continuous representational autoregressive modeling.
\end{itemize}
\vspace{-6pt}
\section{Preliminaries}
\subsection{Variational Autoencoder (VAE)}
Unlike traditional autoencoders, VAE employs latent variables and variational inference to model data distributions explicitly. Specifically, given an observed sequential data \(\mathbf{X} \), VAE introduces continuous latent variables \( \mathbf{Z}^c = \left( \mathbf{z}_t^c \in \mathbb{R}^{d_z^c} \right)_{t=1}^T \) to encode patterns, where \({d_z^c}\) is the dimension of latent variables chosen as a hyperparameter. The generative model (Decoder) \( p_\theta \left( \mathbf{X} \mid \mathbf{Z}^c \right) \) parameterized by \( \theta \) defines the likelihood of observed data conditioned on latent variables, and the inference model (Encoder) \(q_\phi \left( \mathbf{Z}^c \mid \mathbf{X} \right) \) parameterized by \( \phi \) approximates the true posterior distribution.
The joint distribution of observed and latent variables is given by 
\(p_\theta \left( \mathbf{X}, \mathbf{Z}^c \right) = p \left( \mathbf{Z}^c \right) p_\theta \left( \mathbf{X} \mid \mathbf{Z}^c \right) \) 
where the prior distribution is typically assumed as a standard Gaussian distribution \(p \left( \mathbf{Z}^c \right) = \mathcal{N} \left( 0, \mathbf{I} \right) \). Due to the intractability of directly computing the posterior \( p_{\theta}(\mathbf{Z}^c|\mathbf{X}) \), VAE employs the evidence lower bound (ELBO) for optimization:
\begin{equation}
\resizebox{.9\hsize}{!}{$
\log p_{\theta}(\mathbf{X}) \geq 
\underbrace{\mathbb{E}_{q_{\phi}(\mathbf{z}^{c}|\mathbf{x})}\left[\log p_{\theta}(\mathbf{X}|\mathbf{Z}^{c})\right] - D_{KL}\left(q_{\phi}(\mathbf{Z}^{c}|\mathbf{X})||p(\mathbf{Z}^{c})\right)}_{\mathcal{O}_{\text{ELBO}}} $}.
\end{equation}
Maximizing $\mathcal{O}_{\text{ELBO}}$ effectively trains the encoder-decoder architecture, encouraging the inferred latent representation $\mathbf{Z}^c$ to summarize the sequential data compactly.
\vspace{-6pt}
\subsection{Speech Language Model}
Speech language models aim to model speech sequences through latent speech representations, commonly known as speech tokens. 
Following recent works, a general framework for token-based speech modeling typically consists of three components: a speech tokenizer, an autoregressive model, and a decoder. 
Specifically, given the raw speech input $\mathbf{X}$, a speech tokenizer maps $\mathbf{X}$ into a sequence of discrete semantic tokens $\mathbf{Z}^{d} = (z^{d}_t \in \mathcal{N}_k)_{t=1}^{T}$, where $\mathcal{N}_k = \{1,2,\dots,k\} $ denotes a finite vocabulary of speech units. 
The implicit distribution learned by the pretrained tokenizer can be represented as $p(\mathbf{Z}^{d}|\mathbf{X})$.
Next, the autoregressive model parameterized by $ \psi $ models temporal dependencies of the discrete token sequences as 
\vspace{-4pt}
\begin{equation}
    p_{\psi}(\mathbf{Z}^{d}) = \prod_{t=1}^{T} p_{\psi}(z^{d}_t | \mathbf{Z}^{d}_{1:t-1}).
\end{equation}
Finally, a decoder parameterized by $\theta$ reconstructs the original speech $\mathbf{X}$ from the discrete speech tokens $\mathbf{Z}^{d}$, through the conditional distribution $p_{\theta}(\mathbf{X}|\mathbf{Z}^{d})$.
However, since discrete tokens primarily capture linguistic information, this approach tends to overlook the rich paralinguistic features inherent in speech signals, and thus, discrete autoregressive models may struggle to integrate continuous paralinguistic nuances.

\begin{figure*}[h!]   
    \centering
    \includegraphics[width=\textwidth]{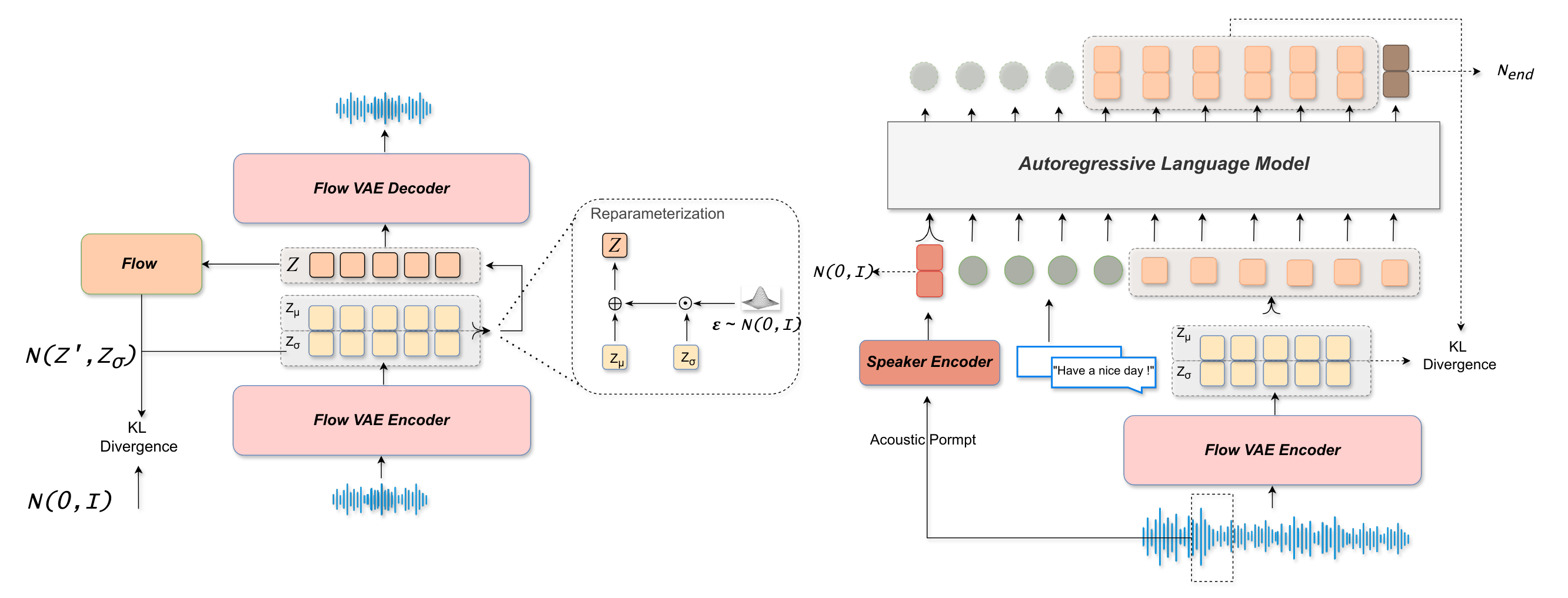}
    \vspace{-6pt}
    \caption{KALL-E Model Architecture: Left: Flow-VAE encoder and decoder for encoding and decoding continuous speech latents. Right: Autoregressive language Model with speaker encoder for text-to-speech generation.} 
    \label{fig:architecture}
    \vspace{-4pt}
\end{figure*}

\vspace{-10pt}
\section{KALL-E}
KALL‑E conceptualizes text‑to‑speech (TTS) as a conditional language‑modeling task framed as next‑distribution prediction. Conventional LLM‑based TTS systems rely on a fixed‑size token dictionary and optimize a cross‑entropy objective to anticipate the next token. This formulation is token‑oriented, aiming to approximate the probability distribution of the forthcoming token. In contrast, KALL‑E learns to predict continuous speech distributions directly, conditioned on textual input.
As illustrated in Figure~\ref{fig:architecture}, the model comprises two principal components: (i) a flow variational auto‑encoder~\cite{minimax-sppech} (Flow‑VAE) and (ii) an autoregressive language model. The Flow‑VAE encodes raw waveforms into a continuous latent space that captures fine‑grained acoustic variation. By integrating a normalizing‑flow module into the conventional VAE, we enhance the expressiveness and diversity of the latent distribution. These continuous latents provide well‑structured training targets for the language model, which predicts them autoregressively from text. Finally, the Flow‑VAE decoder transforms the predicted latents back into the natural‑sounding speech.

\vspace{-6pt}
\subsection{Flow-VAE}

To furnish the autoregressive (AR) language model with continuous speech latents, we first train a variational auto‑encoder (VAE) in an unsupervised manner. 

We augment the vanilla VAE with a normalising flow~\cite{normflow} $f$, which establishes a bijective transformation between a simple prior and a more complex posterior. We note that similar approaches using normalizing flows to enhance prior distributions have also been observed in \cite{vits}~\cite{minimax-sppech} for text-to-speech.
Concretely, we draw $z\sim\mathcal{N}(\mu_{\phi}(x),\sigma_{\phi}(x))$ and obtain $\tilde z = f(z)$. The regularisation objective minimises the Kullback–Leibler divergence between the transformed posterior and the standard normal prior:
\begin{equation}
L_{kl} = D_{KL}(q_\phi(\tilde{z} \mid x) \parallel p(\tilde{z}))),
\end{equation}
\begin{equation}
q_\phi(\tilde{z} \mid x) = N(f(\tilde{z}); \mu_\phi(x), \sigma_\phi(x)) \left| \det \frac{\partial f(\tilde{z})}{\partial \tilde{z}} \right|,
\end{equation}
\begin{equation}
z \sim N(\mu_\phi(x), \sigma_\phi(x)),
\end{equation}
\begin{equation}
p(\tilde{z}) = N(\tilde{z}, 0, I).
\end{equation}
This regularises the encoder to produce Gaussian-distributed latents without forcing them into a unit standard normal, thereby increasing the diversity of the latent space.

As for the detailed architecture of the proposed Flow-VAE, the encoder consists of a stack of down-sampling dilated convolution layers with residual blocks, which effectively capture abstract features in the speech. After encoding, the produced mean and variance are interpreted as the parameters of the learned latent distribution $q_\phi(z \mid x) = N(\mu_\phi(x), \sigma_\phi(x))$. The decoder mirrors the encoder’s architecture but uses transposed convolution layers with residual blocks to up-sample the latent representation $z$ back into the waveform $\hat{x}$. Additionally, we integrate advanced techniques, such as the Snake activation function from BigVGAN~\cite{bigvgan}, to enhance the decoder’s performance.
The Flow‑VAE is optimized with a weighted sum of four losses:
\begin{align}
\mathcal{L}_{\text{Flow-VAE}}
   &= \lambda_{\text{kl}}\,\mathcal{L}_{\text{KL}}
      + \lambda_{\text{recon}}\,\mathcal{L}_{\text{recon}} \notag\\
   &\quad + \lambda_{\text{disc}}\,\mathcal{L}_{\text{disc}}
      + \lambda_{\text{fm}}\,\mathcal{L}_{\text{fm}}.
\end{align}
where $\mathcal{L}_{\mathrm{recon}}$ is an $\ell{1}$ mel‑spectrogram loss, $\mathcal{L}_{\mathrm{disc}}$ combines multi‑period and multi‑resolution discriminator losses, and $\mathcal{L}_{\mathrm{fm}}$ is the feature‑matching loss described in~\cite{featuremap}. Hyperparameters $\lambda$ balance the contribution of each term.

\subsection{Autoregressive Language Model}

With the assistance of Flow-VAE, KALL-E employs a causal transformer decoder as the language model to autoregressively predict the continuous speech distribution. Specifically, the input text tokens are first converted into embeddings by the text embedding layer. Simultaneously, a linear layer projects the sampled speech distribution $z$ into the dimension of the language model. Additionally, a small segment of the speech is randomly extracted and fed into the speaker encoder, which extracts a speaker embedding $S$ and places it at the front of the sequence. The language model, consisting of blocks of multi-head self-attention and feed-forward layers, takes the concatenated speaker, text, and speech embeddings as input, modeling the dependency between semantic and acoustic information.

At each time step $t$, the output of the language model, the output hidden $h_t$, is processed by a linear layer to predict the mean $\mu_t$ and variance $\sigma_t$ of the target speech distribution. These parameters are then used to sample the predicted speech distribution for the subsequent autoregressive (AR) step. This process can be formulated as:
\vspace{-4pt}
\begin{equation}
p(\mu , \sigma\mid S ,text, \theta ) = \prod_{t=1}^{T} p( \mu_t, \sigma_t \mid S ,text, Z_{<t} ,\theta ).
\end{equation}
\begin{equation}
Z_t \sim N(\mu_t, \sigma_t).
\end{equation}
where $\theta$ is the parameter of the language model.

We use the Kullback-Leibler (KL) divergence loss as the training objective for the AR language model. The KL loss has two components: one arises from the difference between the predicted and real speech distributions, and the other is related to stop prediction. Here, we define $N(1,e)$ as the stop distribution. The KL loss is computed as follows:
\vspace{-4pt}
\begin{equation}
\begin{split}
\mathcal{L}_{\text{LM}}
&= \sum_{t=1}^{T}
   D_{\text{KL}}\bigl(q_\phi(z_t \mid X)\;\big\|\;p(Z_t \mid Z_{1:t-1},text,S,\theta)\bigr) \\
&\quad
  +\,\lambda_{end}\,
   D_{\text{KL}}\bigl(N_{end}\;\big\|\;p(e \mid Z,text,S,\theta)\bigr).
\end{split}
\end{equation}
where $N_{end}$ is the pre-defined end distribution, and $\lambda_{end}$ is a hyperparameter that controls its contribution.

\subsection{Speaker Voice Distribution Modeling}

KALL-E without an explicit speaker encoder can handle two extreme cases: (i) reference-guided voice cloning, by prepending one target utterances to the sequence (in-context learning, ICL), the model faithfully mimics that timbre; and (ii) reference-free synthesis, when no audio is supplied, the model hallucinates a plausible but essentially random timbre, which is generally impossible to reproduce later.

\begin{figure}[h!]
    \centering
    \includegraphics[width=\linewidth]{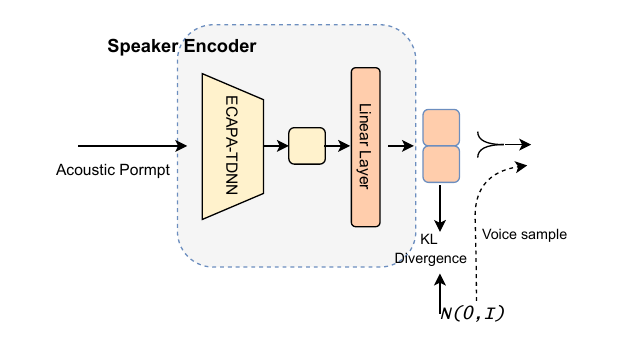}
    \caption{The Architecture of the Speaker Encoder.}
    \label{fig:speaker-encoder}
    \vspace{-10pt}
\end{figure}

To simultaneously achieve high voice diversity and reproducibility, we introduce Speaker Voice Distribution Modeling. As illustrated in Figure \ref{fig:speaker-encoder}, we extract a speaker embedding with ECAPA-TDNN \cite{ecapatdnn} and project it through a linear layer to obtain a latent representation. During training, a KL term regularises the latent toward an isotropic Gaussian prior.

During inference, when a reference utterance is available, we extract its speaker latent $S$ and use it as a condition, yielding deterministic voice cloning. If no reference is provided, we instead draw $\tilde{ S}\!\sim\!\mathcal{N}(\mathbf 0,\mathbf I)$; because the random seed, $\tilde{S}$, can be stored, the resulting timbre can be reproduced exactly, thereby resolving the ``lucky-voice-but-can’t-find-again'' problem. Consequently, KALL-E unifies controllable voice cloning with high-diversity, yet repeatable, random voice synthesis within a single latent framework.

\vspace{-8pt}
\subsection{Test Time Training}
Continuous latents preserve far more fine-grained information than discrete tokens. To exploit this capacity at inference, KALL-E adopts a lightweight test-time training (TTT)~\cite{ttt} procedure that adapts the language model to a new speaker from just one utterance. Given the prompt utterance $S$, we employ the pretrained Flow-VAE to extract the distribution of the prompt speech $S$. This distribution constitutes the supervision signal for TTT.

We draw $N$ latent sequences by repeating the standard reparameterisation trick, and collect them into the TTT dataset:
\begin{equation}
z_i \sim N(\mu_\phi(S), \sigma_\phi(S)).
\end{equation}
\vspace{-8pt}
\begin{equation}
\mathcal D_{\text{TTT}} = \{z_1,z_2 \ldots z_n\}.
\end{equation}
During TTT, we freeze the speaker encoder and fine-tune only the language model. The end–of–sequence distribution is excluded from the loss. Given the fixed transcript, which we assume can always be obtained, and a sampled latent prefix, the model predicts the frame‑level distribution. The loss is the KL divergence between predicted and target distributions, excluding the end‑of‑sequence frame.

This approach effectively alleviates overfitting, which could arise from training with a single sample, and simultaneously enables the model to learn the characteristics of the target distribution more effectively. Consequently, the model better captures detailed information of the target speaker, such as speaking style, accent, and other paralinguistic features. 
\vspace{-6pt}
\section{Experimental Setup}
\subsection{Dataset}
We select the open-source Emilia~\cite{Emilia} as the training dataset. Emilia is a multilingual, diverse, in-the-wild speech dataset designed for large-scale speech generation. In this work, we perform unsupervised training on Flow-VAE using Emilia’s speech data. For the autoregressive language model, we utilize both English and Chinese data from Emilia, totaling approximately 96.7k hours.
Given that the pre-training data may suffer from poor audio quality and inaccurate transcripts, we select a subset with higher audio quality and perform a second round of transcription and cleaning using a different automatic speech recognition (ASR) tool than the one used in the original paper. The cleaned output, combined with our internal data, forms our second-stage fine-tuning dataset, comprising about 3,000 hours, which we also use for small-scale experiments. Additionally, we leverage the ESD~\cite{ESD} dataset to train the model's capability for emotion-aware speech generation.
For evaluation, we follow the procedure of Llasa~\cite{llasa} and use the test-clean subset of LibriSpeech~\cite{Librispeech} to assess the performance of Flow-VAE. The test-clean subset contains 2,620 utterances sampled at 16 kHz. For the review of TTS capabilities, we use the test-zh and test-en subsets of the Seed-TTS~\cite{seedtts} test sets.
\vspace{-6pt}
\subsection{Implementation and Hyperparameters}
For Flow-VAE, the encoder and the flow follow the architecture of Glow-WaveGAN~\cite{glow-wavegan}, while the decoder is based on the settings of BigVGAN~\cite{bigvgan}. The latent dimension of the speech distribution $z$ is set to 512, with a frame rate of 12.5 Hz. For the autoregressive language model, we utilize the open-source LLama3.2-1B-Instruct~\cite{llama} model as our backbone. For the speaker encoder, we randomly initialize an ECAPA-TDNN~\cite{ecapatdnn} module and randomly select 3-second audio segments from the target speech as input, optimizing it together with the autoregressive language model.
We complete the training of Flow-VAE and the autoregressive language model on 8 NVIDIA A100 GPUs. Regarding training hyperparameters, the values of $\lambda_{kl}$, $\lambda_{recon}$, $\lambda_{disc}$, and $\lambda_{fm}$ are set to 32, 1, 1, and 1, respectively. The $\lambda_{end}$ is set to 0.02.
\vspace{-6pt}
\subsection{Comparison Models}
We compare Flow-VAE with several open-source speech tokenizers, including DAC~\cite{dac}, Mimi~\cite{moshi}, Xcodec-2~\cite{llasa}, and Stable Audio VAE~\cite{stable-audio}. Except for Stable Audio VAE, we use the officially released checkpoints. For Stable Audio VAE, we use the same training data to train a 12.5 Hz model with a dimension size of 64 for comparison.
For TTS capabilities, we compare KALL-E with several popular zero-shot TTS models, including Seed-TTS~\cite{seedtts}, FireRedTTS~\cite{fireredtts}, CosyVoice~\cite{cosyvoice}, CosyVoice 2~\cite{cosyvoice2}, and Llasa~\cite {llasa}. Note that the training data used by these models may differ. The model most comparable to KALL-E is Llasa-1B-160k, which has similar parameters and training data, both of which are trained from Llama3.2-1B-Instruct~\cite{llama3}.
For inference efficiency comparison, we also evaluate it against the non-autoregressive language synthesis model F5TTS~\cite{f5tts}. In terms of context awareness, we compare it with Llasa-8B~\cite{llasa}.
\subsection{Evaluation Metrics}

For Flow-VAE evaluation, we adopt the following metrics: (1) Short-Time Objective Intelligibility (STOI); (2) Perceptual Evaluation of Speech Quality (PESQ); (3) a WavLM-based speaker verification model for speaker similarity (SPK SIM)~\cite{wavlm}; and (4) UTMOS~\cite{utmos}. For TTS evaluation, we employ both subjective and objective evaluations. 
Specifically, we use the Mean Opinion Score (MOS) to assess the naturalness of synthetic speech, and the Similarity Mean Opinion Score (SMOS) to evaluate speaker similarity between synthetic and prompt speech. Each evaluation involves at least 10 participants. The rating scale is as follows: 1 (bad), 2 (poor), 3 (fair), 4 (good), and 5 (great), with half-point increments.
Besides, we follow Seed-TTS~\cite{seedtts} and measure Word Error Rate (WER), Character Error Rate (CER), and speaker similarity (SIM). Moreover, to assess inference efficiency, we report giga-floating-point operations per second (GFLOPS), providing a hardware-agnostic metric that helps neutralize variations across test environments.
Additionally, we use Emotion2Vec~\cite{emotion2vec} to assess the emotional category of the generated speech.

\vspace{-6pt}
\section{Experimental Results}
\subsection{VAE Evaluation}

We first quantify the audio reconstruction quality of Flow-VAE and then analyze how the normalizing-flow module influences the latent distribution.

\paragraph{Audio-reconstruction Quality.}
\begin{table*}[htbp]
  \centering
  \footnotesize
  \begin{adjustbox}{max width=\linewidth}
    \begin{tabular}{l c c c c c c c}
      \toprule
      \textbf{Model} & \textbf{Latent Dim} & \textbf{Frame Rate} & \textbf{STOI}↑ & \textbf{PESQ WB}↑ & \textbf{PESQ NB}↑ & \textbf{SPK‑SMI}↑ & \textbf{UTMOS}↑\\
      \midrule
      Ground Truth     &    -  &   - & 1.00  &4.64      & 4.55      &   1.00    &   4.09 \\
      \midrule
      DAC-12 Layer~\cite{dac}     &    -  &  50    & \textbf{0.95}      &  \textbf{4.01}     & \textbf{4.15}      &   \textbf{0.95 }   &  4.00 \\
      DAC-2 Layer~\cite{dac}       &    -  &  50    & 0.73      &  1.13     & 1.40      &   0.32    &  1.29 \\
      Mimi~\cite{moshi}             &    -  &  12.5  &0.91       &2.25       &2.80       &   0.73    &  3.56 \\
      X-codec2~\cite{llasa}         &    -  &  50    & 0.92      &2.43       &3.04       &   0.82    &  \textbf{4.13} \\
      \midrule
      Stable Audio VAE~\cite{stable-audio}     &    64   &   12.5     & 0.96     &3.11      &   3.73   &   0.93   &   4.06 \\
      \multirow{6}{*}{Flow-VAE (Proposed)}
               &    64  &   100     & 0.96     &3.56       & 3.95     &   0.94   &   \textbf{4.07}\\
               &    64  &  12.5     & 0.87     &2.01       & 2.67     &   0.59   &   3.66\\
               &    256 &  12.5     & 0.94     &2.91       & 3.52     &   0.87   &   3.97\\
               &    256 &   25      & 0.96     &3.40       & 3.87     &   0.93   &   4.00\\
               &    512 &  12.5     & 0.96     &3.26       & 3.80     &   0.92   &   3.97\\
               &   1024 &   12.5    & \textbf{0.97}     &\textbf{3.71}       & \textbf{4.05 }    &   \textbf{0.95 }  &   3.99\\
      \bottomrule
    \end{tabular}
  \end{adjustbox}
  \caption{Audio reconstruction results for discrete tokenisers and continuous VAEs on test-clean subset of LibriSpeech. }
  \vspace{-6pt}
  \label{tab:audio_metrics}
  \vspace{-10pt}
\end{table*}
As shown in Table~\ref {tab:audio_metrics}, discrete tokenizers typically require high frame rates to minimize information loss. When DAC is compressed from 12 to 2 layers, PESQ-WB plummets from 4.01 to 1.13 and STOI from 0.95 to 0.73. Even Mimi, which still stacks eight layers at the same 12.5 Hz base rate, remains far below the 12.5 Hz VAE.
In contrast, Flow-VAE can maintain high-quality speech signal reconstruction at low frame rates, outperforming most discrete tokenizers. 
Under the matching 64-dim / 12.5 Hz setting, Stable-Audio VAE outperforms Flow-VAE because its KL penalty on the latent Gaussian prior is almost zero, nearly reducing the model to a plain auto-encoder. 
Flow-VAE deliberately keeps the KL weight at 32, preserving a well-shaped latent space at the cost of reconstruction, an investment that will pay off for generation. 
Though this gap confirms that a stronger Gaussian constraint indeed limits representational capacity, it is important to note that both the frame rate and the latent dimension affect the reconstruction performance of VAE. The reconstruction quality improves as the latent dimension and the frame rate increase, which enlarges the information capacity of latent variables. By increasing the latent dimension, we achieve a high-quality reconstruction while maintaining a low frame rate.

\begin{figure}[h!]
    \centering
    \includegraphics[width=\linewidth]{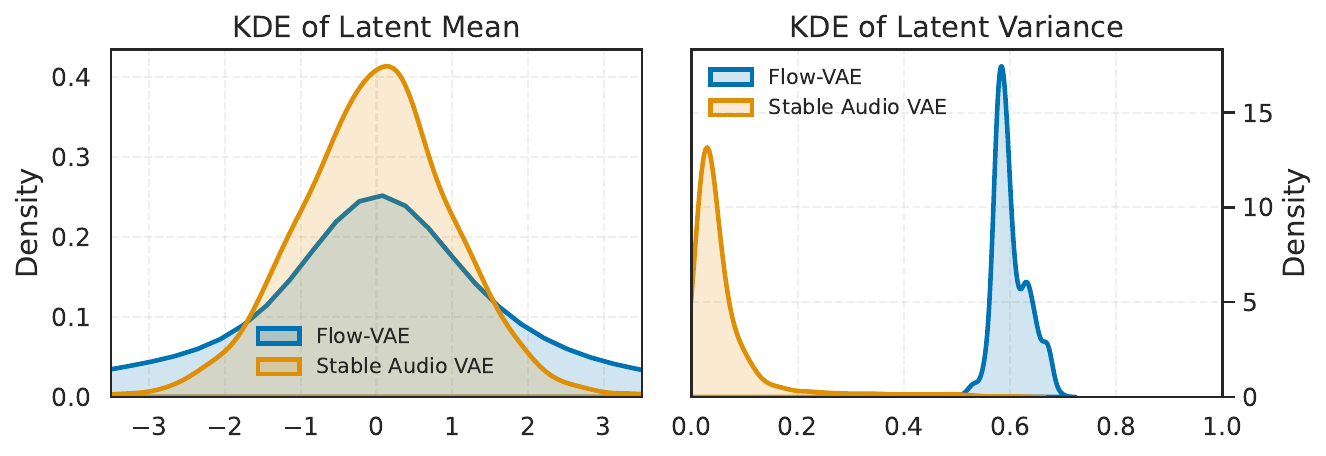}
    \caption{Kernel density estimates(KDE) of the latent representations.}
    \vspace{-4pt}
    \label{fig:kde_of_latent}
\end{figure}

\paragraph{Latent Distribution.}
Unlike conventional VAEs, Flow-VAE does not impose a standard normal distribution on the latent variables, which enhances the diversity of speech distributions. We train both the Stable Audio VAE and Flow-VAE using the same data, frame rate, and latent dimension. We then plot the mean and variance distributions of the extracted audio features. The results are shown in the figure~\ref{fig:kde_of_latent}.

The mean distribution range of Flow-VAE is broader than that of Stable Audio VAE. This indicates that Flow-VAE can map the audio signal to a larger range, which aids the model in distinguishing between different frames. Additionally, we observe that the variance distribution of Flow-VAE significantly differs from that of Stable Audio VAE. Flow-VAE tends to exhibit a larger variance, which improves the robustness of language modeling. This increased variance allows VAEs to tolerate large prediction deviations around the mean. Moreover, this also increases the probability of overlap between different distributions. When predicted points fall into the overlapping regions, there is a higher likelihood of diverse explanations for those points. We speculate that this is also the reason why we can generate diverse speech.
\vspace{-8pt}
\subsection{TTS Evaluation}
In this section we compare KALL‑E with several state‑of‑the‑art LM‑based text‑to‑speech (TTS) systems on the Seed test set~\cite{seedtts}. We selected 12.5 Hz, dim size 512 Flow-VAE to train our KALL-E.
\paragraph{Voice Cloning.}
Table~\ref{tab:cer_wer_similarity} reports character error rate (CER) for Chinese, word error rate (WER) for English, and speaker similarity (SIM). As can be seen from the table, although KALL-E is trained with the smallest amount of data, it achieves the best decoding accuracy on both languages: 0.96 CER on test-zh and 1.94 WER on test-en. The KALL-E (TTT), which incorporates test-time training, shows a better performance in terms of SIM, while maintaining similar CER and WER compared to the original KALL-E.
Interestingly, several discrete-token systems report SIM scores higher than KALL-E. 
Because these systems do condition on the target-speaker reference audio when decoding tokens into waveforms, we hypothesize that this extra conditioning inflates objective similarity scores and may also influence the perceived naturalness of the synthesized speech. We investigate this possibility in the following subjective test.
\begin{table}[h!]
  \centering
  \begin{adjustbox}{max width=\linewidth}
  \begin{tabular}{l cc cc}
    \toprule
    \multirow{2}{*}{\textbf{Model}} &
      \multicolumn{2}{c}{\textbf{test‑zh}} &
      \multicolumn{2}{c}{\textbf{test‑en}} \\
    \cmidrule(lr){2-3}\cmidrule(lr){4-5}
      & \textbf{CER $\downarrow$} & \textbf{SIM $\uparrow$} & \textbf{WER $\downarrow$} & \textbf{SIM $\uparrow$} \\
    \midrule
    Human                & 1.26 & 0.755 & 2.14 & 0.734 \\
    \midrule
    Seed‑TTS             & 1.12 & \textbf{0.796} & 2.25 & \textbf{0.762}\\
    FireRedTTS           & 1.51 & 0.635 & 3.82 & 0.460 \\
    CosyVoice            & 3.63 & 0.723 & 4.29 & 0.609 \\
    CosyVoice 2          & 1.45 & 0.748 & 2.57 & 0.652 \\
    \midrule
    Llasa-1B-160k        & 2.22 & 0.658 & 3.60 & 0.563 \\
    KALL-E               & \textbf{0.96} & 0.646 & 1.94 & 0.568 \\
    \midrule
    KALL-E (TTT)         & 1.02 & 0.698 & \textbf{1.90} & 0.611 \\
    \bottomrule
  \end{tabular}
  \end{adjustbox}
  \caption{Recognition error rates and speaker similarity on the Chinese
    (\textit{test‑zh}) and English (\textit{test‑en}) sets.
    Lower CER/WER and higher SIM indicate better performance.}
    \vspace{-10pt}
  \label{tab:cer_wer_similarity}
\end{table}

We conduct a subjective evaluation on Seed-TTS test sets, assessing the naturalness and speaker similarity between synthetic and reference speech.
Fifteen native listeners participate in the evaluation via a web interface. Table~\ref{tab:subject_evaluation} reports the
mean MOS for naturalness and the SMOS for speaker similarity.
\begin{table}[htbp]
  \centering
  \begin{adjustbox}{max width=\linewidth}
  \begin{tabular}{l c r}
    \toprule
      \textbf{Model} & \textbf{MOS$\uparrow$} & \textbf{SMOS$\uparrow$}  \\
      \midrule
      KALL-E      & \textbf{4.17 $\pm$ 0.08 }     & 3.93 $\pm$ 0.15 \\
      Llasa 1B    & 3.92 $\pm$ 0.12      & 3.85 $\pm$ 0.08 \\
      Llasa 8B    & 3.97 $\pm$ 0.10      & 3.92 $\pm$ 0.11 \\
      CosyVoice~2 & 4.03 $\pm$ 0.16      & 3.92 $\pm$ 0.08 \\
      F5TTS       & 3.93 $\pm$ 0.07      & 3.87 $\pm$ 0.12 \\
      \bottomrule
    \end{tabular}
  \end{adjustbox}
  \caption{Subjective Evaluation on Seed-TTS test sets: Naturalness (MOS) and Speaker Similarity (SMOS).}
  \vspace{-10pt}
  \label{tab:subject_evaluation}
\end{table}
Listeners rate KALL-E highest on naturalness, significantly outperforming Llasa-1B and on par with the much larger Llasa-8B. SMOS differences are within one interval width for the top systems, indicating comparable voice fidelity.

\paragraph{Inference Efficiency.}
To quantify generation efficiency, we compute the floating‑point operations (GFLOPs) required to synthesise 10s of speech (Table~\ref{tab:infer_speed}). A lower codec frame rate (12.5 Hz) offsets KALL‑E’s large parameter count, making it faster than all baselines and showing the high inference efficiency.
\begin{table}[htbp]
  \centering
  \begin{adjustbox}{max width=\linewidth}
  \begin{tabular}{l c c r}
    \toprule
      \textbf{Model} & \textbf{Parameter} & \textbf{Frame Rate} & \textbf{Flops (Gflpos)$\downarrow$} \\
      \midrule
      KALL-E      & 1B   & 12.5 & \textbf{7\,947.48} \\
      Llasa 1B    & 1B   & 50   & 122\,170.02 \\
      CosyVoice~2 & 500M & 25   & 12\,095.08 \\
      F5TTS       & 300M & 93.75& 11\,546.98 \\
      \bottomrule
    \end{tabular}
  \end{adjustbox}
  \caption{Comparison of 10s audio inference speeds for different models. In order to eliminate the impact of the test environment on the final results, we calculated the number of floating-point operations of different models. }
  \vspace{-10pt}
  \label{tab:infer_speed}
\end{table}
\vspace{-4pt}
\paragraph{Context Awareness.}

\begin{figure}[htbp]
    \centering
    \includegraphics[width=\linewidth]{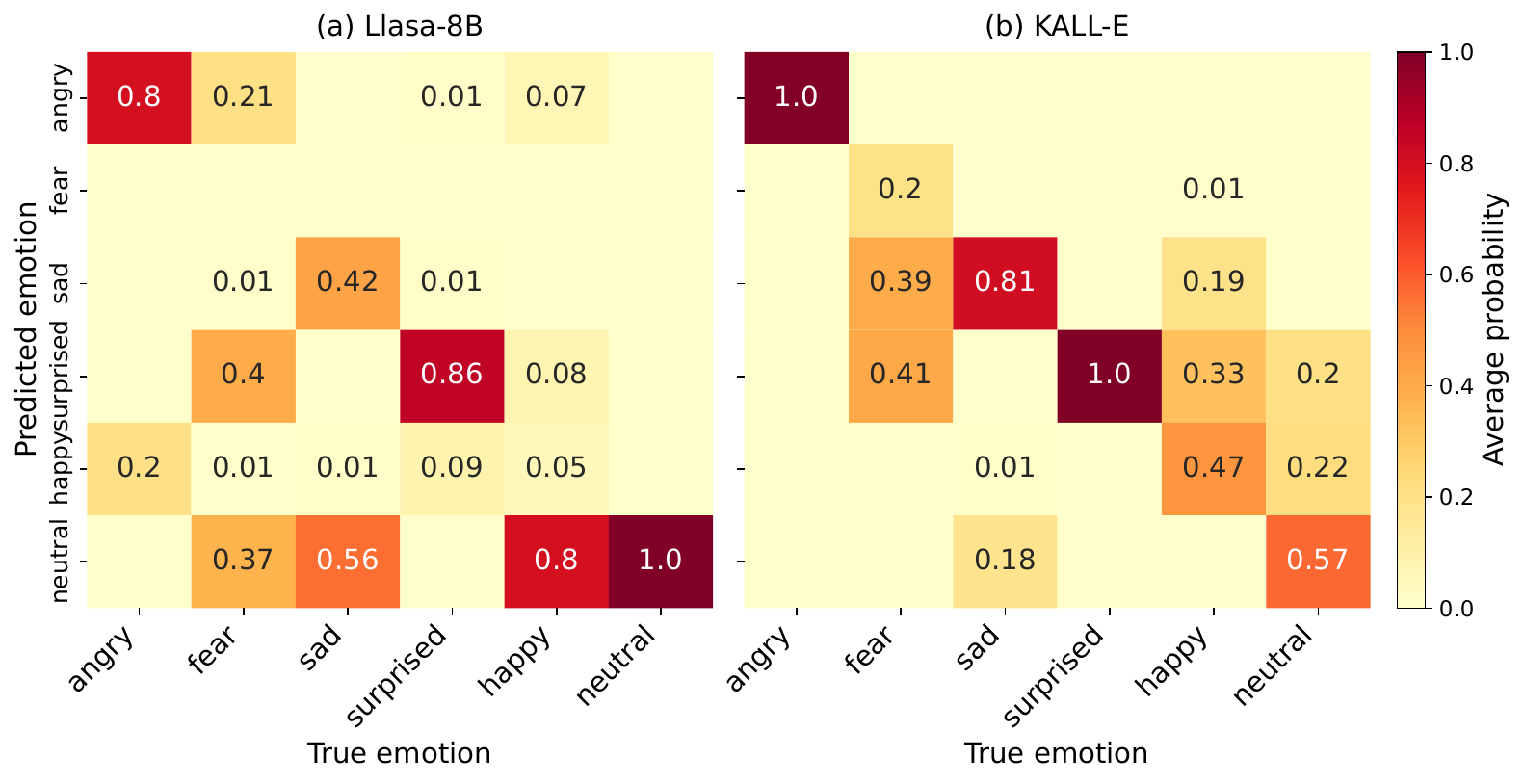}
    \caption{Heatmap of emotion recognition confusion matrices for synthetic speech (LLaSA-8B vs. KALL-E).}
    \vspace{-10pt}
    \label{fig:emotion_heat}
    \vspace{-4pt}
\end{figure}

To further assess the context awareness of KALL‑E, we leverage  ChatGPT to create emotional texts and then synthesize speech only using these texts as input. Finally, we use Emotion2Vev to recognize the emotion and draw the confusion matrix. As shown in Figure~\ref{fig:emotion_heat}, compared to Llasa-8B, KALL‑E can infer the more appropriate emotion solely from the input text without any emotional speech prompt.  We attribute this ability to two aspects. On the one hand, we started training from the text-pretrained LM, and the model itself has a certain understanding of semantics. On the other hand, most of the paralinguistic information, such as emotions, is retained in the continuous representation, which is conducive to the model learning the mapping from text to emotions.

\subsection{Ablation Study}
\paragraph{Flow-VAE Ablation}
In this section, we compare the impact of different Flow-VAE variants on LM modeling. We also compare Flow-VAE and Stable Audio VAE of the same dimension to further investigate the impact of introducing flow on LM modeling. Experiments are conducted on small-scale data.

\begin{figure}[h!]
    \centering
    \includegraphics[width=\linewidth]{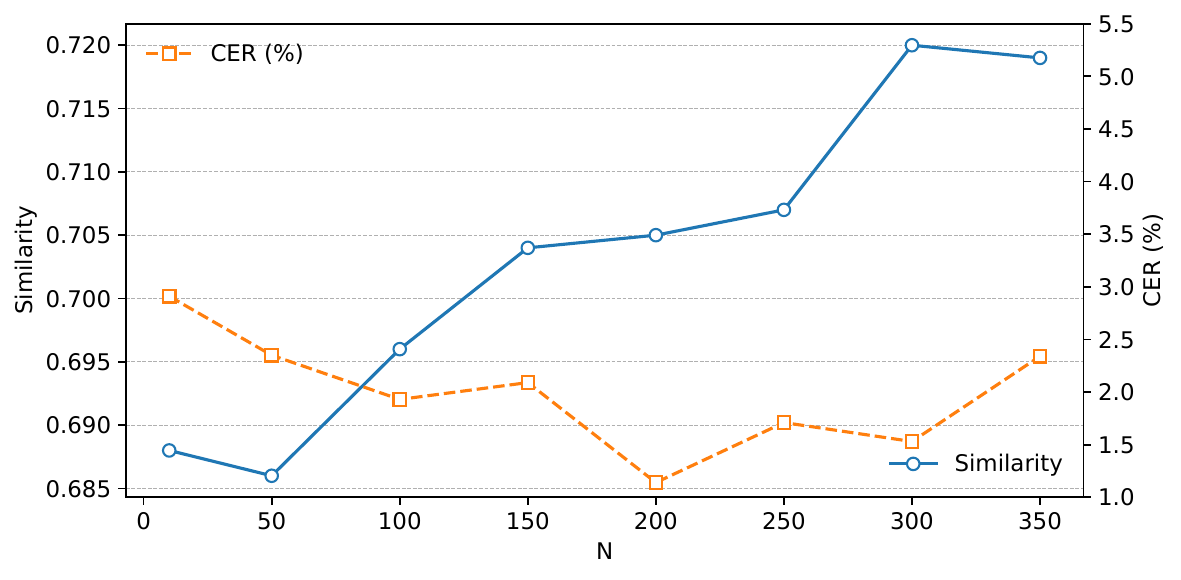}
    \vspace{-10pt}
    \caption{Effect of TTT Training Set Size (N) on speaker similarity and character error rate.}
    \vspace{-8pt}
    \label{fig:TTT}
\end{figure}

The table~\ref{tab:latent_conf} shows that reducing the latent dimension has little impact on WER but a greater impact on the final SIM metrics. This suggests that when encoding information, VAE discards detailed acoustic information first. Furthermore, when replacing Flow-VAE with Stable Audio VAE, model performance significantly degraded. This suggests that our constructed Flow-VAE is more suitable for LM modeling.
\begin{table}[htbp]
  \centering
  \begin{adjustbox}{max width=\linewidth}
  \begin{tabular}{l c c c r}
    \toprule
      \textbf{Model} & \textbf{Latent Dim} & \textbf{Frame Rate} & \textbf{CER $\downarrow$} & \textbf{SIM $\uparrow$} \\
      \midrule
      Flow-VAE    & 64      & 12.5      &  6.15 & 0.435  \\
      Flow-VAE    & 256     & 12.5      &  4.91 & 0.438  \\
      Flow-VAE    & 256     & 25        &  7.30 & 0.401  \\
      Flow-VAE    & 512     & 12.5      &  \textbf{2.79} & \textbf{0.495}  \\
      Flow-VAE    & 1024    & 12.5      &  19.69 & 0.357  \\
      \midrule
      Stable Audio VAE & 64      & 12.5      &  40.09 & 0.117  \\
      \bottomrule
    \end{tabular}
  \end{adjustbox}
  \caption{Comparison of Flow-VAE and Stable Audio VAE Encoder Configurations: Impact on Language Model Character Error Rate and Similarity}
  \vspace{-10pt}
  \label{tab:latent_conf}
\end{table}
\vspace{-6pt}
\paragraph{Test Time Training (TTT) Ablation}
In this section, we verify the ultimate improvement in model performance achieved by TTT. To enhance the comparison, we primarily selected test items from the seed Chinese test set with a non-zero CER index, combined with a few randomly selected items, for a total of 200 items as the test set.  During adaptation, we kept the learning rate fixed at $1\!\times\!10^{-6}$ and varied the size of the \textsc{TTT} training subset $N$.  Figure~\ref{fig:TTT} plots the resulting speaker similarity (left axis) and CER (right axis).
Similarity increases almost monotonically with $N$, suggesting that most speaker-specific information can be captured with TTT. CER falls sharply around $N\!=\!200$, after which it climbs again. We attribute the rebound to mild over-fitting: the model begins to mimic disfluencies or idiosyncratic pronunciations present in the reference. These findings indicate that TTT is sufficient to enhance both perceived identity and intelligibility.

\vspace{-10pt}
\section{Conclusion}
In this paper, we propose KALL-E, a novel model that predicts speech distribution for speech synthesis.
By integrating the flow module into the traditional VAE architecture, KALL-E introduces a continuous representation extracted via Flow-VAE, which is better suited for autoregressive modeling. This continuous representation allows KALL-E to achieve efficient and fast speech synthesis at a low frame rate. 
Additionally, KALL-E demonstrates strong context awareness, enabling the model to infer emotions from text without any additional emotional speech prompts. The incorporation of TTT further enhances the model’s performance.
Experimental results demonstrate the effectiveness of the proposed model and its ability to generate natural, expressive, and context-aware speech.

\bibliography{aaai2026}
\newpage




\end{document}